\documentclass[fleqn,twoside]{article}
\usepackage{espcrc2}

\usepackage{graphicx}
\usepackage[figuresright]{rotating}

\newcommand{\AmS}{{\protect\the\textfont2
  A\kern-.1667em\lower.5ex\hbox{M}\kern-.125emS}}

\title{Detecting chiral singularities in lattice QCD at strong coupling}

\author{Costas G. Strouthos 
 \address{Division of Science and Engineering, 
        Frederick Institute of Technology, 
        Nicosia 1303, Cyprus} 
}

\begin{document}

\begin{abstract}
We study the difficulties associated with detecting chiral
singularities in strongly coupled lattice
QCD at fixed nonzero tempearture. We show that the behavior
of the chiral condensate, the pion mass and the pion decay constant, for
small masses, are all consistent with the predictions of Chiral Perturbation Theory (ChPT).
However, the values of the quark masses that we need to demonstrate
this are much smaller than those being used in dynamical QCD simulations.

\vspace{1pc}
\end{abstract}

\maketitle

\section{Introduction}
Given the difficulties associated with understanding chiral singularities
in a realistic QCD calculation, we explore this subject in strong coupling lattice
QCD with one staggered fermion at finite temperature.
The model undergoes a second order transition which belongs to the $O(2)$ universality
class \cite{chandra}. At fixed $T<T_c$, within the $O(2)$ scaling window, the long distance physics
of our model is described by a $3d$ $O(2)$ continuum scalar field theory in its broken phase.
At this temperature there is a range of quark masses where the light pions are describable
by a $3d$ continuum chiral perturbation theory (ChPT).
The effective action can be written in terms of $\vec{S}$, an $O(2)$ vector field
with the constraint $\vec{S}\cdot\vec{S}=1$. At the lowest order this is given
by
\begin{equation}
S_{\rm eff} = \int d^3 x
\Big[\ \frac{F^2}{2} \ \partial_\mu \vec{S} \cdot \partial_\mu \vec{S} +
 \Sigma \ \vec{h} \cdot \vec{S} \Big],
\label{chlag}
\end{equation}
where $\vec{h}$ is the magnetic field. $F$ is the pion decay constant and $\Sigma$ is the chiral
condensate. In three dimensions $F^2$ has dimensions of mass. 

We connect our model to ChPT and minimize lattice artifacts by choosing $T$ and the quark mass $m$ such
that $ 1 \ll \xi \ll 1/M_{\pi}$, where $M_{\pi}$ is the pion mass.
In this region we expect that $\Sigma$, $M_{\pi}$ and $F^2$ satisfy the expansion
\begin{equation}
\langle O \rangle =  z_0 + z_1\sqrt{m} + z_2 m + z_3 m \sqrt{m} + ...,
\end{equation}
where the $\sqrt{m}$ behavior is the power-like singularity arising due to the infrared
pion physics \cite{walla}.

The partition function of lattice QCD with one staggered fermion interacting with 
$U(3)$ gauge fields can be mapped into a monomer-dimer system in the strong coupling limit \cite{rossi}. 
In this study we use a very efficient algorithm discovered recently to solve the model in the 
chiral limit \cite{adams}.
In this work we choose $L_x=L_y=L$. One can study the thermodynamics of the model
by working on asymmetric and anisotropic lattices with
$L_t \ll L$ and allowing the temporal staggered fermion phase factor to vary continuously \cite{strou}. 

In order to study chiral physics we focus on the chiral condensate $\Sigma_m$, the chiral susceptibility $\chi$, 
the helicity 
modulus $Y$ and the pion mass $M_{\pi}$.
As discussed in \cite{hasen}, one can define
the pion decay constant at a quark mass $m$ to be
$F_m^2 \equiv \lim_{L\rightarrow \infty} Y$. For $m=0$ we then obtain
$F = \lim_{m\rightarrow 0} F_m$, the pion decay constant introduced
in Eq.(\ref{chlag}). We can also define $\Sigma = \lim_{m\rightarrow 0} \Sigma_m$, 
where $\Sigma_m$ is the infinite volume chiral condensate.

\section{Results}
For massless quarks we fit $F^2$ vs. $T$ to $F^2 = C(T_c-T)^{\nu}$  for $T<T_c$.
In the fit we fix $T_c=7.47739(3)$ \cite{chandra}  and $\nu=0.67155(27)$ \cite{campo}.
For $7.05 \leq T \leq 7.42$ we get $C=0.2217(1)$ with $\chi^2/DOF=1.03$. Including $T=7.0$ in the fit makes
$\chi^2/DOF$ jump to $2.0$. At $T$ close to $T_c$, $\xi$ is very large and it is difficult to satisfy
$\frac{1}{M_{\pi}} \gg \xi$. On the other hand when $\xi \sim 1$ lattice artifacts become important.
We estimate that $T=7.0$ is at the edge of the $O(2)$ scaling window.
All the results discussed below were obtained at $T=7.0$.
We fix $L_t=4$ and vary spatial volume from $L=8$ to $L=96$ for
$0 \leq m \leq 0.025$.

The first four terms in the chiral expansion of $F_m$ in $3d$
$O(N)$ ChPT are given by \cite{hasen}:
\begin{equation}
F_m = F[1 + a_1 \sqrt{m} + a_2 m + a_3 m\sqrt{m}],
\label{fpi}
\end{equation}
where $a_1 \propto (N-2)$. Since in our case $N=2$ we expect $a_1=0$.
\begin{figure}[t!]
\centering
\includegraphics[scale=0.28]{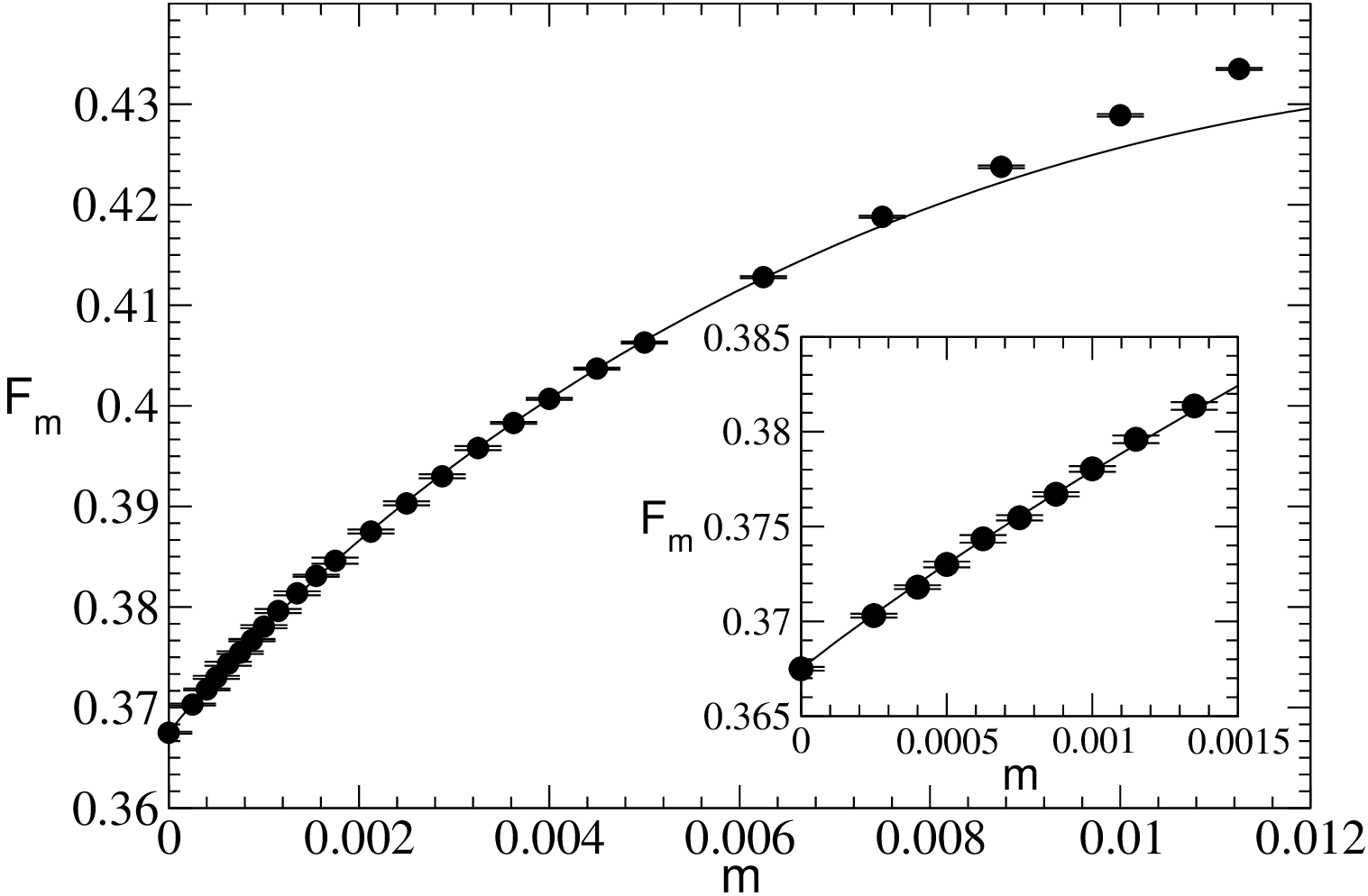}
\caption{Plot of $F_m$ vs. $m$. The solid line represents the function 
$F_m=0.36747[1+34.3m-185m\sqrt{m}]$.}
\label{fig:fig1}
\end{figure}
We fit data in the range $0 \leq m \leq 0.00625$ (see Fig.\ref{fig:fig1}). We find $F=0.3674(1)$, $a_1=0.01(2)$,
$a_2=34.0(7)$ and $a_3 =-182(6)$ with $\chi^2$/DOF$=1.2$.
The prediction of ChPT that $a_1=0$ is in excellent
agreement with our results.

ChPT predicts that the first four terms in the
chiral expansion of $\Sigma_m$ are given by
\begin{equation}
\Sigma_m = \Sigma [1 + b_1 \sqrt{m} + b_2 m + b_3 m\sqrt{m}].
\label{cc}
\end{equation}
As shown in \cite{hasen} $b_1 \propto (N-1)$, implying that $b_1 \neq 0$ in our case. 
At $m=0$ we compute $\Sigma$ by using the finite size scaling formula for $\chi$
given by \cite{hasen}
\begin{equation}
\chi = \frac{1}{N}\Sigma^2 L^3\Big[ 1 + 0.226 (N-1) \frac{1}{F^2 L} +
\frac{\alpha}{L^2} +...\Big],
\label{chptchi}
\end{equation}
where $\alpha$ depends on the higher order low energy constants of the
chiral Lagrangian.
When $F$ is fixed to $0.3675$ obtained earlier, our
results for $16 \leq L \leq 144$ fit very well to this formula. We find
$\Sigma = 2.2648(10)$, $\alpha=4.6(3)$ with $\chi^2$/DOF$=0.87$.
\begin{figure}[t!]
\centering
\includegraphics[scale=0.28]{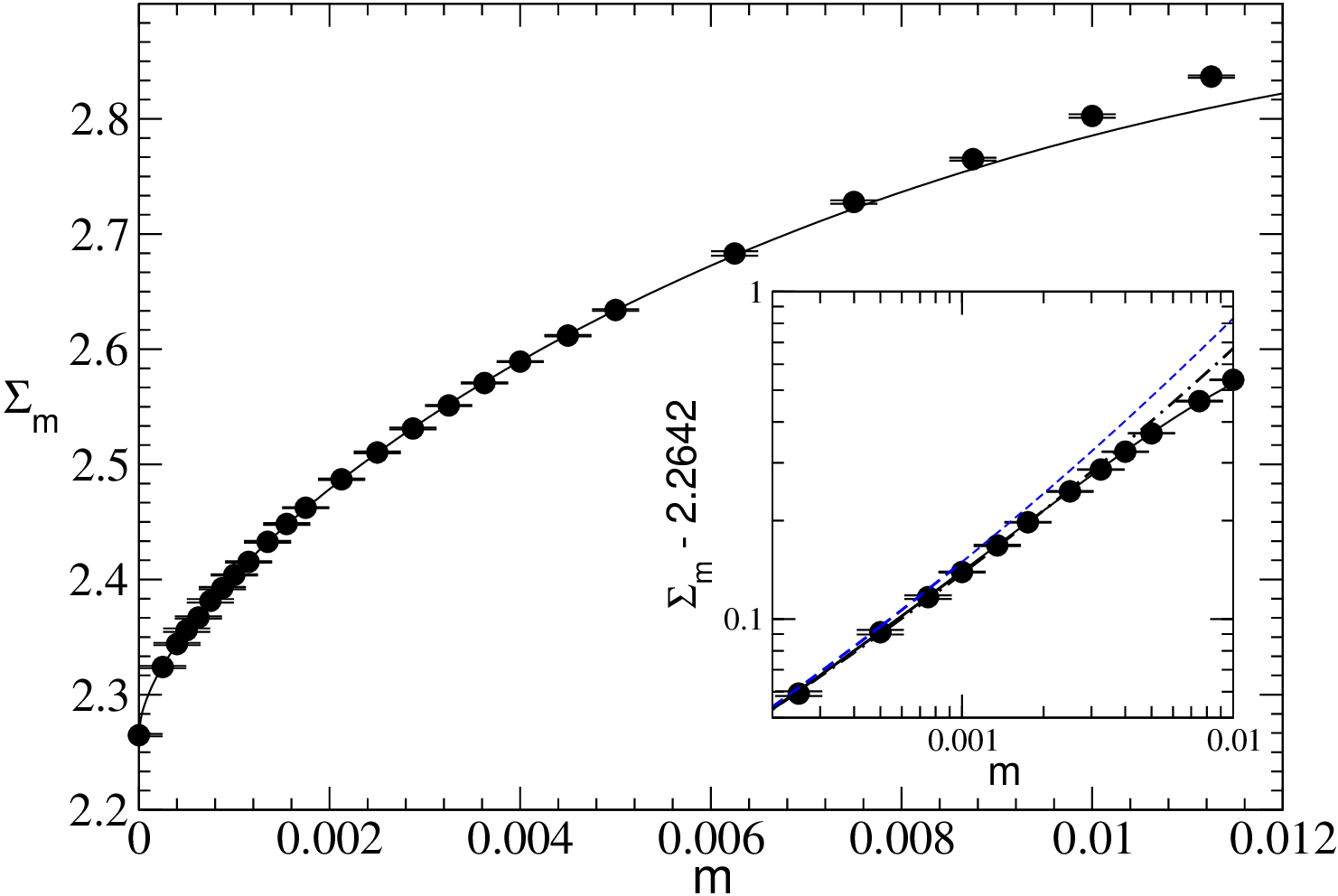}
\caption{Plot of $\Sigma_m$ vs. $m$. The solid line represents the function 
$\Sigma_m=2.2642[1+1.36\sqrt{m}+23m-135m\sqrt{m}]$. The dashed line in the inset 
represents $\Sigma_m -2.2642 = 2.2643[1.44\sqrt{m} + 15.7 m]$. For
clarity some of the data points are not shown in the inset.}
\label{fig:fig2}
\end{figure}
Fitting our data to Eq.(\ref{cc}) in the region $0 \leq m \leq 0.00625$
gives $\Sigma=2.2642(10)$,
$b_1=1.36(4)$, $b_2 = 23(1)$, $b_3 = -135(10)$ with a $\chi^2$/DOF$=1.1$.
We note that we cannot find a mass range within our results
in which we can find a good fit when we fix $b_2=b_3=0$. However,
in the range $0 \leq m \leq 0.00175$ we can set $b_3=0$
to obtain $\Sigma=2.2643(10)$, $b_1=1.44(3)$ and
$b_2=15.2(7)$ with a $\chi^2$/DOF$=1$. Note that $b_2$
changes by about 30\% when this different fitting procedure is
used, while $b_1$ is more stable. The data and the fits are shown in 
Fig.\ref{fig:fig2}.

The first four terms in the chiral behavior of the pion mass are
predicted to be of the form
\begin{equation}
{\tiny M_\pi = \sqrt{\frac{\Sigma}{F^2}}\sqrt{m}
\Big[1 + c_1 \sqrt{m} + c_2 m + c_3 m\sqrt{m}\Big]}.
\label{mpi}
\end{equation}
\begin{figure}[t!]
\centering
\includegraphics[scale=0.28]{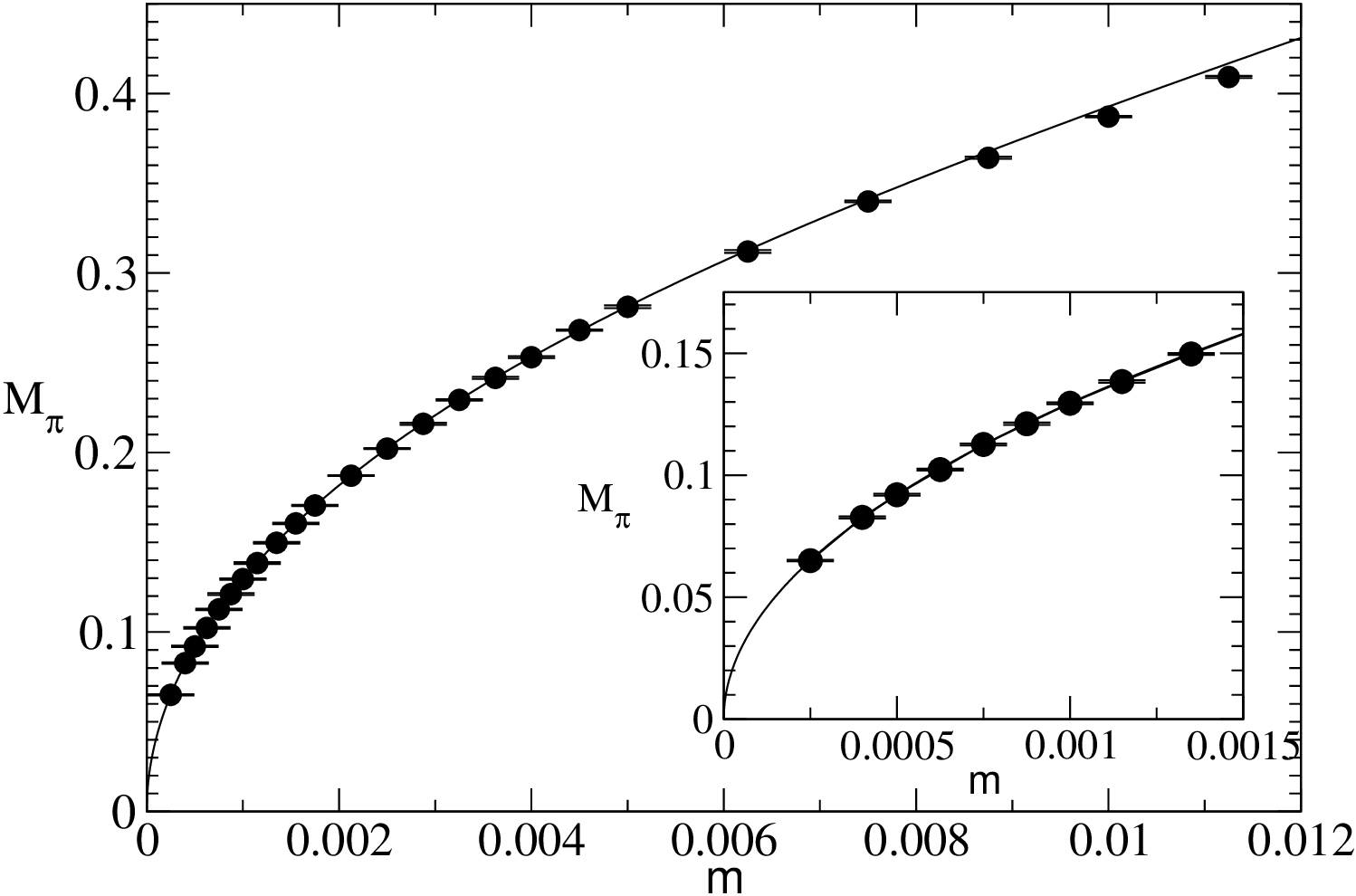}
\caption{Plot of $M_\pi$ vs. $m$. The solid line
represents the function
$M_\pi = 4.095 \sqrt{m}[1+0.68 \sqrt{m} - 25.8 m + 149 m \sqrt{m}]$.}
\label{fig:fig3}
\end{figure}
Further, the chiral Ward identities imply that $c_1 = b_1/2 - a_1$.
Fitting our data to Eq.(\ref{mpi}) after fixing $\Sigma=2.2648$
and $F=0.3675$ obtained above, we find $c_1 = 0.55(16)$,
$c_2=-21(6)$ and $c_3 =110(50)$ with $\chi^2/$DOF$=0.5$. We see that
our data is consistent with the relation $c_1=b_1/2 - a_1$
although the error in $c_1$ is large. Fixing $c_1=b_1/2=0.68$ obtained
from fitting the chiral condensate,
yields $c_2=-25.8(8)$ and $c_3=149(14)$ without changing the
quality of the fit. This fit along with our data for $M_{\pi}$
are shown in Fig.\ref{fig:fig3}.

\section{Discussion}
It is expected that the natural expansion parameter for ChPT
in three dimensions is $x\equiv M_\pi/(4\pi F^2)$.
Let us find
the values of $x$ where the chiral expansion up to a certain power
of $x$ is sufficient to describe the data reasonably.
\begin{figure}[t!]
\centering
\includegraphics[scale=0.29]{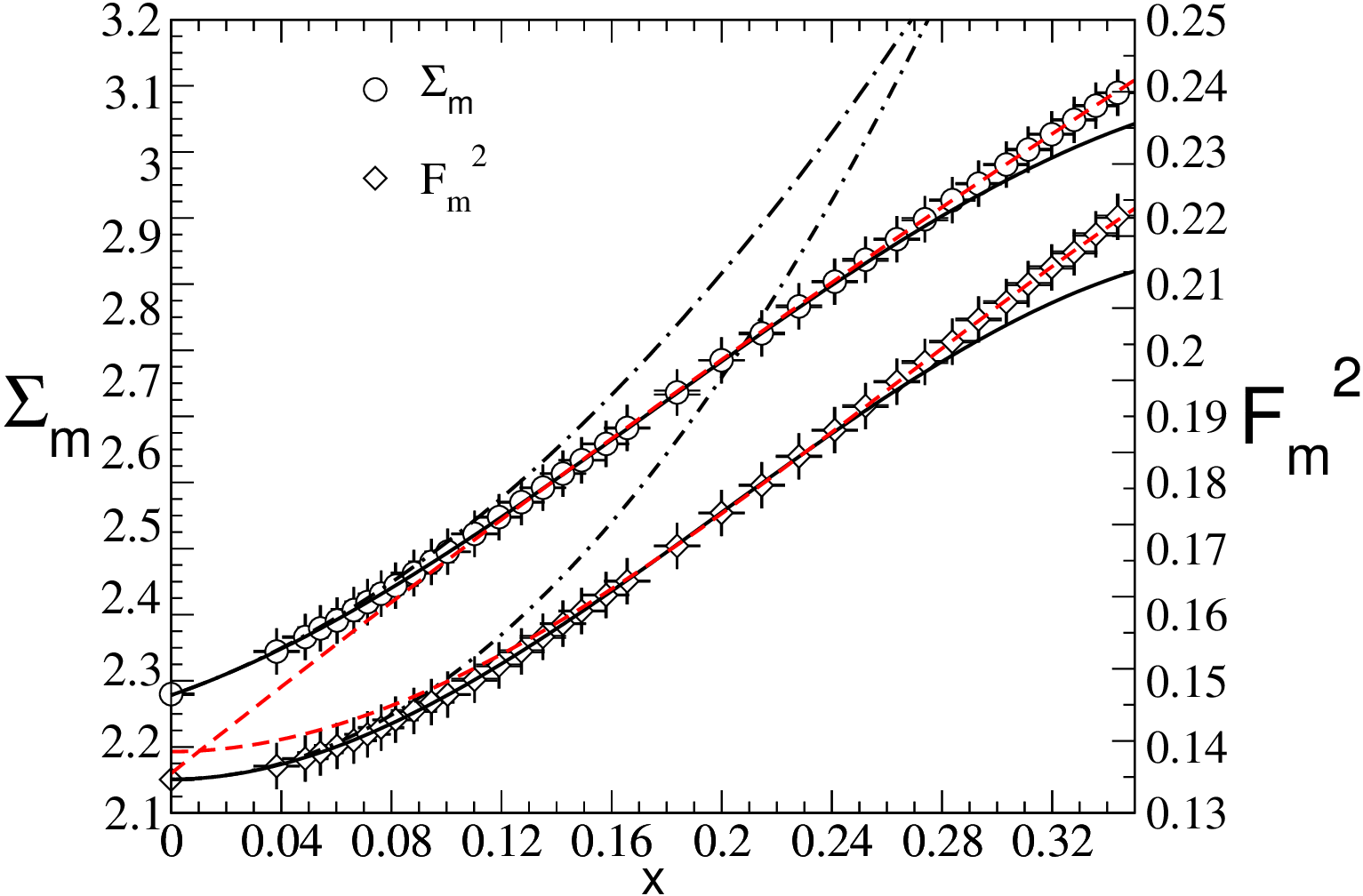}
\caption{Plot of $F^2_m$ and $\Sigma_m$ vs.
$x\equiv M_\pi/(4\pi F^2)$. The solid lines and dot-dashed lines are
fits to two different orders of chiral expansion discussed in the text.
The dashed line represent fits using only the data with $x \geq 0.18$.}
\label{fig:fig4}
\end{figure}
The solid lines in Fig.\ref{fig:fig4} are the best fits to the form $z_0 + z_1 x + z_2 x^2
+ z_3 x^3$, while the dot-dashed lines are the best fits to a smaller
range in $x$ with $z_3$ set to $0$. As the graph indicates the relative
error in $\Sigma_m$ due to the absence of the $z_3$ term is smaller from
the error in $F_m^2$ at a given value of $x$. In order to determine $F_m^2$
within say $5\%$ we need the $z_3$ term even at $x \sim 0.15$. Thus, we
conclude that in our model the chiral expansion converges rather poorly,
especially for the pion decay constant.
It is striking that almost all previous reliable unquenched lattice QCD
simulations with staggered fermions have used $m \geq 0.005$ (in lattice
units), irrespective of whether the study was of QCD thermodynamics or
the QCD vacuum, whether the studies were done at strong couplings or
weaker couplings.
The value $m\sim 0.005$ corresponds
to $x\sim 0.18$ in our model. We can ask how the chiral extrapolations
will look if we use the data for $x \geq 0.18$ to fit to the expected
chiral form.  In Fig.\ref{fig:fig4} we show the best fits for large masses using 
the dashed lines. As expected in order to use chiral
extrapolations reliably it is important to know the range of their
validity.

\section*{Acknowledgements}
This work was done in collaboration with Shailesh Chandrasekharan.

\end{document}